# Ultracompact quantum splitter of degenerate photon pairs

**Jiakun He, Bryn A. Bell, Alvaro Casas-Bedoya, Yanbing Zhang, Chunle Xiong,[*] and Benjamin J. Eggleton**

*Center for Ultrahigh bandwidth Devices for Optical Systems (CUDOS), Institute of Photonics and Optical Science (IPOS), School of Physics, University of Sydney, NSW 2006, Australia*
*\*Corresponding author: chunle@physics.usyd.edu.au*

Compiled 25th June 2015

**Integrated sources of indistinguishable photons have attracted a lot of attention because of their applications in quantum communication and optical quantum computing. Here, we demonstrate an ultra-compact quantum splitter for degenerate single photons based on a monolithic chip incorporating Sagnac loop and a micro-ring resonator with a footprint of 0.011 mm², generating and deterministically splitting indistinguishable photon pairs using time-reversed Hong-Ou-Mandel interference. The ring resonator provides enhanced photon generation rate, and the Sagnac loop ensures the photons travel through equal path lengths and interfere with the correct phase to enable the reversed HOM effect to take place. In the experiment, we observed a HOM dip visibility of 94.5 ± 3.3 %, indicating the photons generated by the degenerate single photon source are in a suitable state for further integration with other components for quantum applications, such as controlled-NOT gates.**

Integrated photonic circuits have emerged as a promising approach to quantum technologies such as quantum cryptography, which offers provably secure communication [1], and quantum information processing, where for certain computational tasks an exponential speed-up is predicted compared to classical information processing [2,3]. Sources of single photons and entangled pairs of photons are essential to many schemes for long-distance quantum communications and optical quantum computing, and in particular pairs of photons that are degenerate in wavelength and indistinguishable from one another have proven useful in proof-of-principle demonstrations of photonic logic gates [4], quantum algorithms [3], entanglement generation [5] and quantum simulations [6]. Degenerate pairs have traditionally been generated using spontaneous parametric down conversion in bulk-nonlinear crystals [7]. However, on-chip photonic circuits with integrated sources of indistinguishable photon pairs are required in order to make this technology scalable.

Recently, pair generation has been demonstrated in silicon waveguides using spontaneous four-wave mixing (SFWM), potentially allowing integration of photon sources and circuits with electronic elements [8–10]. While non-degenerate photons can be split by wavelength de-multiplexing, this is not possible for degenerate photon pairs, since the two photons share the same properties in all degrees of freedom: spatial, wavelength and polarization [11–13]. Splitting these photons probabilistically with a 50:50 directional coupler means they are no longer useful for some tasks; for example a Hong-Ou-Mandel (HOM) dip with probabilistically split photons is limited to a visibility of 50%. An elegant solution is to deterministically split the photons using time-reversed HOM interference. So far, this has been demonstrated in fiber Sagnac loops [14,15] and with planar waveguides in a Mach-Zehnder interferometer (MZI) [16,17]. In those schemes, a degenerate pair is created in one of two waveguides which are then combined at a 50:50 coupler in-phase such that the pair splits into two single photons, in a reversal of the usual HOM experiment where two single photons meet at a coupler and bunch into a pair.

In this paper, we demonstrate degenerate pair generation and deterministic splitting using an integrated Sagnac loop for the first time. Additionally, the actual generation of photon pairs take place in a ring resonator coupled to the Sagnac loop, enhancing the generation efficiency, as well as ensuring the photons are created in well-defined cavity modes and hence have suitable properties for quantum experiments [18,19]. This circuit is ultra-compact, with 50 times smaller area than the silicon MZI based circuit in Ref. [16], and is intrinsically stable, requiring zero tunable elements. For previous implementations in fiber loops and on-chip MZIs, active tuning of the phase to split the photon pairs were required. In our Sagnac configuration, however, the generated photon pairs is automatically in phase of each other. We believe that this compactness, its compatibility with current CMOS technology and the absence of phase-tuning will lead to convenient scaling to more complicated circuits requiring multiple sources.

Figure 1(a) shows a schematic of the working principle. A Sagnac loop is formed by a silicon multimode interference (MMI) coupler, whose two outputs are connected. A silicon ring resonator is coupled to the loop serving as the nonlinear device for degenerate photon pair generation, via the SFWM process, as illustrated in Fig. 1(b). When the ring is pumped at two resonant frequencies, $\omega_{p1}$ and $\omega_{p2}$, one photon at each frequency is annihilated to generate two correlated photons at the central resonant frequency $\omega_{s,i} = (\omega_{p1} + \omega_{p2})/2$ according to energy conservation. When the ring is pumped simultaneously in both clockwise and counter-clockwise directions, in the power regime that multi-pair generation is unlikely, the output photon pair state from the ring resonator is:

$$|\Psi\rangle_{ring} = \frac{|2\rangle_{cw}|0\rangle_{ccw} + e^{i\delta\varphi}|0\rangle_{cw}|2\rangle_{ccw}}{\sqrt{2}} \quad (1)$$

the $\delta\varphi$ term in the equation indicates the relative phase of the photons generated between the clockwise (cw) and counter-clockwise (ccw) directions. The probability to generate photon pairs in one direction should be exactly the same as the probability to generate photon pairs in the other direction. The corresponding output state after the MMI is given by:

$$|\Psi\rangle_{MMI} = \frac{1 - e^{i\delta\varphi}}{2}|\Psi\rangle_{2002} + \frac{i(1 + e^{i\delta\varphi})}{2}|\Psi\rangle_{11} \quad (2)$$

where:

$$|\Psi\rangle_{2002} = \frac{|2\rangle_A|0\rangle_B - |0\rangle_A|2\rangle_B}{\sqrt{2}} \quad (3)$$

$$|\Psi\rangle_{11} = |1\rangle_A|1\rangle_B \quad (4)$$

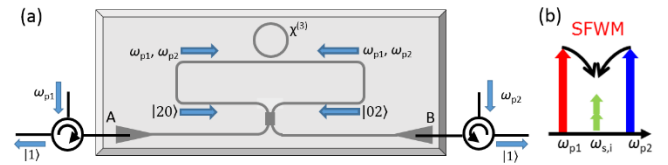

Fig. 1. (a) Schematic of the working principle: waveguides A and B, connected to two grating couplers, reach the same side of a multimode interference coupler (MMI), whose two output ports of the MMI coupler are connected, forming a Sagnac loop; a silicon ring resonator is coupled to the Sagnac loop in order to enhance the photon-pair generation rate. (b) Schematic of degenerate photon-pair generation via SFWM. Two pump photons at different wavelengths are annihilated to create a pair of degenerate single photons.

A and B denote the two output ports of the MMI coupler for the photon pairs. The beam-splitter transformation of the MMI gives a $\pi/2$ phase shift between the two directions for each pump, so that if the two pumps are injected into the same input, the total phase $\delta\varphi = \pi$, and the output state is $|\Psi\rangle_{2002}$. However, by injecting the pumps from two separate input ports, we ensured a total relative phase of zero ($\delta\varphi = 0$), with the $\pi/2$ phase shifts of the two pumps cancelling out. As a result, when the photon pairs generated from the ring resonator reach the MMI coupler from the Sagnac loop side, they will split deterministically, leaving $|\Psi\rangle_{MMI} = |\Psi\rangle_{11}$.

Figure 2 shows the experimental set-up. The two pump waves were generated by spectrally slicing the broadband (30 nm bandwidth) output of a mode-locked erbium doped fiber laser (38.6 MHz) using a spectral pulse shaper (SPS – Finisar Waveshaper). This allowed two short pump pulses to be synchronized temporally. The resulting pump waves were flat top shaped with a spectral bandwidth of 0.1 nm centered at $\lambda_{p1}$ =1540.8 nm and $\lambda_{p2}$ =1558.1 nm, corresponding to two resonances of the ring resonator separated by four free spectral ranges. The 0.1 nm bandwidth for both pumps are approximately twice the full-width-half-maximum (FWHM) of the ring's resonances, ensuring the resonances are covered, and that the cavity defines the spectra of the photons, not the pumps. After the spectral splicing, a low noise Erbium doped fiber amplifier (EDFA, Pritel) was employed to amplify the pump power. A fiber 50:50 coupler was used to split the pumps into two separate fibers, in order to inject pump 1 into port A, and pump 2 into port B of the sample. Tunable filters (Santac and Dicon) were then used to only let one of the two pump waves pass, in order to prevent leakage from the undesired pump and ASE noise. Polarization controllers were used to match the pumps' polarizations to the waveguide's TE mode, as the on-chip grating couplers used to couple light to and from the sample and waveguides were designed to support the TE mode only. A circulator was used at each input to separate outgoing light for collection. The gray area in Figure 2 shows an SEM image of the silicon on insulator sample. It was fabricated at IMEC through ePIXfab using 193 nm deep UV lithography [20] and has a device footprint of $207 \times 54$ μm $= 0.011$ mm$^2$. The nanowires and the ring resonator have cross-section dimensions of $450 \times 220$ nm. This waveguide has an anomalous dispersion in the telecom C-band and hence has a high SFMW efficiency. The effective interaction length in the resonator is much longer than the silicon bus waveguide, so SFMW from inside the ring is dominant compared to that the bus waveguide. The on chip ring resonator has a radius of 21 um, its Q-factor was measured to be $4 \times 10^4$, with the free spectral range (FSR) of 4.33 nm near 1550 nm. During the experiment, the on-chip powers of the two pump waves were matched to maximize the degenerate SFWM efficiency [21].

Light coupled off-chip, after having been separated from the incoming pump light by the circulators, was sent to two band-pass filters (BPFs) tuned to the degenerate photons' wavelength of 1549.4 nm, to provide a total pump isolation of 100 dB. After the spectral filtering, the photons coming out from the two sides were directed to two separate superconducting single photon detectors (SSPDs, Single Quantum, ~10% detection efficiency with 100 Hz dark count, polarization sensitive). Finally, coincidence measurements were conducted using a time interval analyzer (TIA). Degenerate photons generated and successfully split by the Sagnac and micro-ring circuit should emerge from separate outputs and register as coincidences between the two detectors.

Throughout the experiment, the TIA was used to obtain the raw coincidence rate ($C_{raw}$) and accidental coincidence rate ($A$) [10]. The true coincidence count was then calculated as $C = C_{raw} - A$. We plot the count rate versus total coupled average power in Fig. 3. Detected photon pair coincidences (red circles) far exceed the accidental coincidences (green triangles), demonstrating correlated photon-pair generation. A quadratic fit of the true coincidences is shown. The quadratic dependence of the coincidence rate on power is a signature for SFWM processes. A roll off of the coincidence rate when power is above 100 μW is due to the two-photon absorption (TPA) inside the ring resonator and corresponding free-carrier absorption (FCA) [22]. We also plot the measured coincidence-to-accidental ratio (CAR) on the right y-axis, at different pump powers. In our experiment, we recorded a maximum CAR of 235 ± 21 at an average pair detection rate of 1.4 Hz. At a greater detection rate of 3.9 Hz, the CAR was measured to be 61 ± 3.

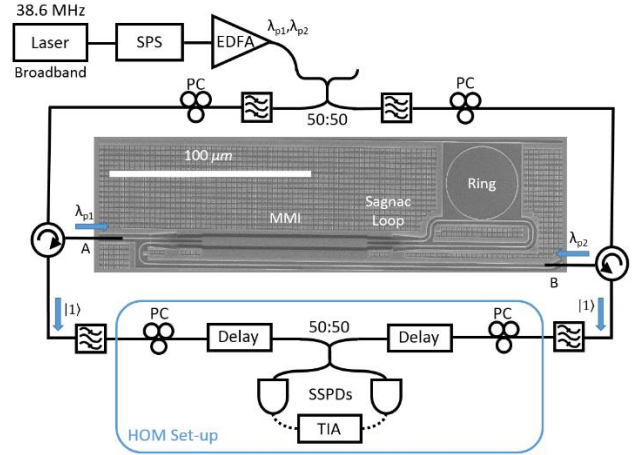

Fig. 2. Experimental setup. A broadband mode-locked laser was spectrally sliced into two synchronized pumps using a spectral pulse shaper (SPS) and subsequently amplified by an Erbium doped fiber amplifier (EDFA); a fiber 50:50 coupler was used to split the pumps into two; Band pass filters (BPFs) were used to ensure only pump $\lambda_1$ was injected to port A and only pump $\lambda_2$ was injected to port B; polarization controllers (PCs) matched the pumps to the TE mode required by the waveguide. Right before injecting the pump waves into the sample, a circulator on each side was inserted for collection of generated single photons. After coupling the photons out from the sample, BPFs were used to reject the pumps. The photons were then either directly connected to superconducting single photon detectors (SSPDs); or connected through a HOM set-up consisting of tunable delay lines, PCs and a fiber 50:50 coupler (indicated inside blue lines); the gray area inside the figure shows a SEM image of the device.

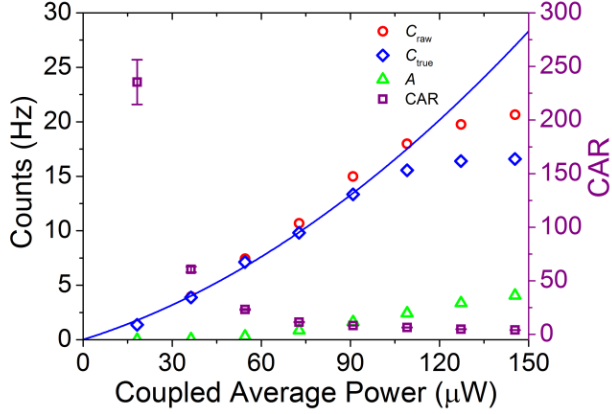

Fig. 3. Counts (left axis) and CAR (right axis, purple squares, with error bars) as a function of coupled average power. Measured raw coincidences (red circles, $C_{raw}$), accidental coincidences (green triangles, $A$), and true coincidences (blue diamonds, $C = C_{raw} - A$) are shown in the plot. The blue line is a quadratic fit of $C_{true}$, when coupled $P_{avg} < 90$ μW. With coupled power higher than 100 μW the counts are rolling off from the fit due to two-photon absorption (TPA) and free-carrier absorption (FCA).

Detection of correlated pairs between the two outputs does not necessarily indicate that time-reversed HOM interference took place, because even if the photons behave according to classical statistics at the MMI, up to 50% of the pairs can emerge from separate outputs. In order to test the time-reversed HOM interference, we measured the splitting ratio of the photon pairs. Having first measured the coincidence rate between the two outputs, we added a 50:50 coupler after the spectral filtering at each output from the sample, and connected the two outputs of the 50:50 couplers to SSPDs. Any photon pairs emerging from the same output would probabilistically split at these couplers, yielding a 50% probability to measure a coincidence count. In this way, by comparing the number of pairs coming out from the same port to the number of photon pairs coming out from two separate outputs of the sample, the splitting ratio can be calculated. At a typical power level, 89.9 ± 6.7 % of the pairs were emerging from separate outputs, indicating that the time-reversed HOM effect was successfully splitting the photons, as this is a significantly larger proportion than the 50% expected from probabilistic splitting.

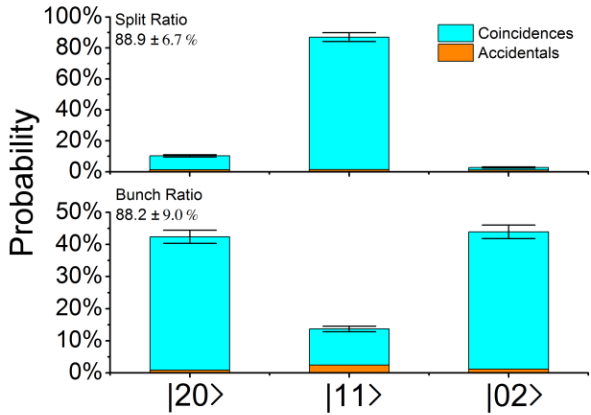

Fig. 4. Count rate for split ($|11\rangle$) and bunched ($|20\rangle$ and $|02\rangle$) pairs. Top: when two pumps were injected separately into the two ports of the sample, splitting ratio of 88.2 ± 9.0 % after subtracting noise is achieved; Bottom: when two pump were injected together into one of the ports of the sample, bunching ratio of 85.8 ± 6.1 % after subtracting noise is achieved.

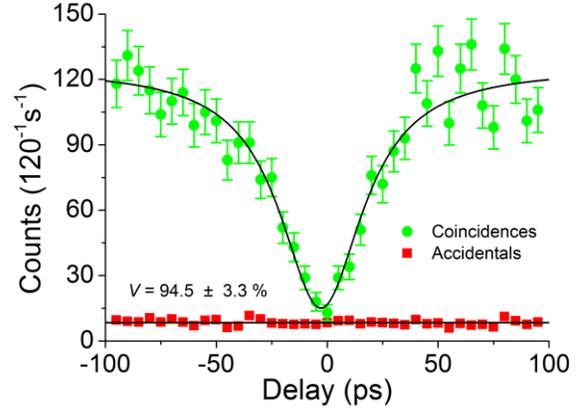

Fig. 5. Hong-Ou-Mandel experimental results. A visibility of 94.5 ± 3.3 % is observed after subtracting accidental coincidences. Green circles indicate raw coincidences, error bars are Poissonian; red squares indicate accidental coincidences; black lines are Lorentzian fit for the raw coincidences and linear fit for the accidental coincidences.

To further elaborate our result on the photon splitting ratio, we changed our pump configuration so that both pump waves were injected to the same input of the Sagnac loop, setting the relative phase of the photons' state in the ring to $\delta\varphi = \pi$. As a result, when the photons generated from the ring resonator arrive at the MMI coupler, they should always bunch and emerge from the same output port ($|\Psi\rangle_{MMI} = |\Psi\rangle_{2002}$). Under this new pump configuration, we measured the splitting ratio. As seen in Fig. 4 (bottom), 43.5 ± 4.4 % of the pairs came out together from port A of the Sagnac loop and another 44.7 ± 4.5 % of the pairs came out together from port B of the Sagnac loop, together, 88.2 ± 9.0 % of the photon pairs are emitted in bunched state, $|\Psi\rangle_{2002}$. Note this is a NOON state with $N=2$, and has applications in quantum metrology for enhanced measurement of a phase [23].

The deviation from the theoretical 100% photon splitting or bunching ratio is mostly caused by the coherent back scattering of light in the silicon ring resonator into the opposite direction, which has been investigated classically [24,25]. Here, the significant coupling between the two directions in the ring resonator causes the experiment to differ from the theoretical wave-function given in equation (1), in the case where generated photons are being back scattered. This is mostly due to the side wall roughness of the ring resonators. By adopting improved waveguide fabrication technologies, the surface roughness can be reduced, which would directly reduce the back scattering, and hence a higher splitting or bunching ratio may be achieved.

A high splitting ratio does not necessarily imply that the photons are indistinguishable from one another, though this is expected if they are created in the same resonant mode of the ring. To test the indistinguishability of the split photon pairs, a HOM dip setup was added as shown inside the blue lines in Figure 2. Instead of connecting directly to the SSPDs after the spectral filters on each side, the photons were first coupled to fiber tunable optical delay lines (General Photonics, 1 ps step) and polarization controllers, and then meet at a single-mode fiber based 50:50 coupler, where HOM interference took place. The two outputs of the last 50:50 coupler were connected to two SSPDs. Changing the relative time delay ($\delta t$) with the tunable delay line varies the temporal overlap between the two photons. When the two photons were made to arrive simultaneously ($\delta t = 0$), a reduction in coincidence count rate was observed because the photons were bunching and were not expected to emerge at the separate outputs of the 50:50 coupler. The coincidence results are plotted in Figure 5, as a function of δt. The plot shows a HOM dip with raw visibility 88.1 ± 3.1% and corrected visibility 94.5 ± 3.3%, once the flat background of accidentals is subtracted. Both exceed the threshold of 50% possible for classical light sources or probabilistically split photon pairs, as well as demonstrating the indistinguishability of the photons. The noise present at relative delays beyond 50 ps is due to the coupling fluctuations over a long measurement time. This clearly non-classical behavior demonstrates the suitability of the compact Sagnac and ring resonator source design for use in lager quantum photonic circuits.

In conclusion, we demonstrated, for the first time, an ultra-compact quantum splitter for degenerate single photons based on a monolithic chip incorporating Sagnac loop and a micro-ring resonator. The device is compatible with current CMOS technology and has a footprint 50 times smaller than the previous published results, generating and deterministically splitting degenerate photon pairs using time-reversed HOM interference. We observed an off-chip HOM dip visibility of 94.5 ± 3.3 %, indicating this degenerate pair photon source is in a suitable state for further integration with other components, for further quantum applications, such as controlled-NOT gates [4].

This experiment is funded by Australian Research Council (ARC) Centre of Excellence Program (CE110001018), ARC Laureate Fellowship (L120100029), ARC Discovery Early Career Researcher Award (DE120100226). We thank Dr. Alex Clark for useful discussion.

**REFERENCES**

1. N. Gisin and R. Thew, "Quantum communication," Nature Photon. **1**, 165 (2007).
2. L. Grover, "Fixed-Point Quantum Search," Phys. Rev. Lett. **95**, 150501 (2005).
3. A. Politi, J. Matthews, and J. O'Brien, "Shor's quantum factoring algorithm on a photonic chip," Science **325**, 1221 (2009).
4. A. Politi, M. J. Cryan, J. G. Rarity, S. Yu, and J. L. O'Brien, "Silica-on-silicon waveguide quantum circuits," Science **320**, 646 (2008).
5. D. Bouwmeester, J.-W. Pan, M. Daniell, H. Weinfurter, and A. Zeilinger, "Observation of three-photon Greenberger-Horne-Zeilinger entanglement," Phys. Rev. Lett. **82**, 1345 (1999).
6. A. Aspuru-Guzik and P. Walther, "Photonic quantum simulators," Nat. Phys. **8**, 285 (2012).
7. D. Höckel, L. Koch, and O. Benson, "Direct measurement of heralded single-photon statistics from a parametric down-conversion source," Phys. Rev. A **83**, 013802 (2011).
8. J. Sharping, K. Lee, M. Foster, A. Turner, B. Schmidt, M. Lipson, A. Gaeta, and P. Kumar, "Generation of correlated photons in nanoscale silicon waveguides," Opt. Express **14**, 12388 (2006).
9. K. Harada, H. Takesue, H. Fukuda, T. Tsuchizawa, T. Watanabe, K. Yamada, Y. Tokura, and S. Itabashi, "Frequency and polarization characteristics of correlated photon-pair generation using a silicon wire waveguide," IEEE J. Sel. Top. Quantum Electron. **16**, 325 (2010).
10. C. Xiong, C. Monat, A. S. Clark, C. Grillet, G. D. Marshall, M. J. Steel, J. Li, L. O'Faolain, T. Krauss, J. Rarity, and B. Eggleton, "Slow-light enhanced correlated photon pair generation in a silicon photonic crystal waveguide," Opt. Lett. **36**, 3413 (2011).
11. J. He, A. S. Clark, M. J. Collins, J. Li, T. F. Krauss, B. J. Eggleton, and C. Xiong, "Degenerate photon-pair generation in an ultracompact silicon photonic crystal waveguide," Opt. Lett. **39**, 3575 (2014).
12. Y. Guo, W. Zhang, S. Dong, Y. Huang, and J. Peng, "Telecom-band degenerate-frequency photon pair generation in silicon microring cavities," Opt. Lett. **39**, 2526 (2014).
13. J. Fan, A. Migdall, and L. Wang, "Efficient generation of correlated photon pairs in a microstructure fiber," Opt. Lett. **30**, 3368 (2005).
14. J. Chen, K. Lee, C. Liang, and P. Kumar, "Fiber-based telecom-band degenerate-frequency source of entangled photon pairs," Opt. Lett. **31**, 2798 (2006).
15. L. Yang, F. Sun, N. Zhao, and X. Li, "Generation of frequency degenerate twin photons in pulse pumped fiber optical parametric amplifiers: Influence of background noise," Opt. Express **22**, 2553 (2014).
16. J. W. Silverstone, D. Bonneau, K. Ohira, N. Suzuki, H. Yoshida, N. Iizuka, M. Ezaki, C. M. Natarajan, M. G. Tanner, R. H. Hadfield, V. Zwiller, G. D. Marshall, J. G. Rarity, J. L. O'Brien, and M. G. Thompson, "On-chip quantum interference between silicon photon-pair sources," Nature Photon. **8**, 104 (2013).
17. H. Jin, F. M. Liu, P. Xu, J. L. Xia, M. L. Zhong, Y. Yuan, J. W. Zhou, Y. X. Gong, W. Wang, and S. N. Zhu, "On-Chip Generation and Manipulation of Entangled Photons Based on Reconfigurable Lithium-Niobate Waveguide Circuits," Phys. Rev. Lett. **113**, 103601 (2014).
18. S. Clemmen, K. Phan Huy, W. Bogaerts, R. Baets, P. Emplit, and S. Massar, "Continuous wave photon pair generation in silicon-on-insulator waveguides and ring resonators," Opt. Express **17**, 16558 (2009).
19. S. Azzini, D. Grassani, M. J. Strain, M. Sorel, L. G. Helt, J. E. Sipe, M. Liscidini, M. Galli, and D. Bajoni, "Ultra-low power generation of twin photons in a compact silicon ring resonator," Opt. Express **20**, 23100 (2012).
20. S. K. Selvaraja, W. Bogaerts, P. Dumon, D. Van Thourhout, and R. Baets, "Subnanometer Linewidth Uniformity in Silicon Nanophotonic Waveguide Devices Using CMOS Fabrication Technology," IEEE J. Sel. Top. Quantum Electron. **16**, 316 (2010).
21. Q. Lin, F. Yaman, and G. Agrawal, "Photon-pair generation in optical fibers through four-wave mixing: Role of Raman scattering and pump polarization," Phys. Rev. A **75**, 023803 (2007).
22. C. a Husko, A. S. Clark, M. J. Collins, A. De Rossi, S. Combrié, G. Lehoucq, I. H. Rey, T. F. Krauss, C. Xiong, and B. J. Eggleton, "Multi-photon absorption limits to heralded single photon sources," Sci. Rep. **3**, 3087 (2013).
23. V. Giovannetti, S. Lloyd, and L. Maccone, "Advances in quantum metrology," Nature Photon. **5**, 222 (2011).
24. F. Morichetti, A. Canciamilla, M. Martinelli, A. Samarelli, R. M. De La Rue, M. Sorel, and A. Melloni, "Coherent backscattering in optical microring resonators," Appl. Phys. Lett. **96**, 13 (2010).
25. F. Ladouceur and L. Poladian, "Surface roughness and backscattering," Opt. Lett. **21**, 1833 (1996).